\documentclass[nofootinbib]{revtex4}
\usepackage[english]{babel}
\usepackage{array,booktabs}   
\usepackage{array} 
\usepackage{amsmath} 
\usepackage{relsize}
\usepackage{wrapfig}
\usepackage{calc}
\usepackage{sitem}
\usepackage{pdflscape}
\usepackage{color}
\usepackage{float}
\usepackage{graphicx}
\usepackage{amsmath}
\usepackage[nodisplayskipstretch]{setspace}

\usepackage{setspace}
\usepackage{tabularx}

\usepackage{float}
\usepackage{color}
\usepackage{amsmath}
\usepackage{float}
\usepackage{calc}
\usepackage{pdflscape}
\usepackage{color}
\usepackage{float}
\usepackage{graphics}
\usepackage{epsfig}
\usepackage{epstopdf}
\usepackage{amssymb}
\usepackage{graphicx}
\usepackage{epstopdf}
\usepackage{appendix}
\usepackage{romannum}
\usepackage{soul}
\usepackage{color}

\definecolor{blizzardblue}{rgb}{0.67, 0.9, 0.93}
\definecolor{bubblegum}{rgb}{0.99, 0.76, 0.8}
\usepackage[urlcolor=blizzardblue]{hyperref}
\hypersetup{
	colorlinks=false,
	linkcolor=blue,
	filecolor=magenta,      
	urlcolor=bubblegum,
}
\usepackage{relsize}

\usepackage[raggedright]{titlesec}
\begin{document}

	\pagenumbering{arabic}
	\title{Uncertainties in QE and RES events at LBNF due to hadronic production in FSI}
	\author{ Ritu Devi$^{1}$\footnote{E-mail: rituhans4028@gmail.com}, Jaydip Singh$^{2}$\footnote{E-mail: jaydip.singh@gmail.com}, Baba Potukuchi$^{1}$\footnote{E-mail: baba.potukuchi@gmail.com}}
	
	\affiliation{Department Of Physics, University of Jammu, Jammu, India$^{1}$}
	\affiliation{Department Of Physics, University of Lucknow, Lucknow, India.$^{2}$}

	\bigskip
	\begin{abstract}
 \textbf{Abstract}\\
To achieve appropriate interaction rates in recent neutrino oscillation studies, high atomic number nuclear targets were utilized. Because of the nuclear effects in the experimental observable, the utilization of these complicated targets produced systematic uncertainties that needed to be assessed accurately to constrain the discovery. We made an effort to calculate the nuclear effects in the Ar and H targets, which are intended to be employed at the DUNE distant and near detectors, respectively, through our simulation effort. The DUNE flux is peaking around 2.5 GeV and CCRES is the dominant process at this energy. So, this work will be focused only on the CCQE and CCRES interactions and the simulations will be done using two different neutrino event generators. We reported the ratio of the oscillation probability (P(Ar)/P(H) as a function of reconstructed neutrino energy for CCRES channels to quantify the systematic errors in the observables. 
\end{abstract}

\maketitle
\section {Introduction}	\label{sec1}

DUNE will be a long-baseline neutrino oscillation experiment \cite{1,1DUNE,2DUNE,3DUNE}. The important research goal of this experiment is  constraining the CP violation phase in the leptonic sector \cite{4DUNE} and  to determine the parameters governing neutrino oscillation. Another goal of the DUNE project is to determine the mass hierarchy, whether it is  an inverted hierarchy (IH) or a normal hierarchy (NH), which is currently unknown, as well as atmospheric neutrino physics, supernova physics  \cite{1b},  and proton decay physics \cite{1a}. DUNE will be a neutrino and anti-neutrino spectrum ($ \nu_{\mu}(\bar{\nu_{\mu}}), \nu_{e}(\bar{\nu_{e}})$, and $\nu_{\tau}(\bar{\nu_{\tau}})$) beam facility at Fermilab, with a near detector (ND) near the neutrino source at Fermilab and a far detector (FD) at the Sanford Underground Research Facility (SURF) in South Dakota, USA. This will give a 1248-kilometer baseline facility for studying the matter effect. To study the neutrino spectrum, both detectors will use Ar target material, which will enable to overcome numerous systematic errors. The ND will look at the spectra of un-oscillated neutrinos, whereas the FD will look at the spectrum of oscillated neutrinos. DUNE chose to use the PRISM concept in ND, which suggested that two detectors be used, one on-axis and the other traveling across the beam. On a daily basis, the beam stability is monitored using the System for on-Axis Neutrino Detection (SAND). SAND employed a hydrocarbon target and $\nu(\bar\nu)$-H from $CH_{2}$ subtraction, which provides useful information for reducing systematic uncertainty \cite{sand}. The precise energy measurement of each flavor at both detectors is critical for completing the experiment's scientific purpose. The events rate as a function of reconstructed neutrino energy at both the ND and FD detectors, which are used to constrain values of the neutrino oscillation parameters and also proton lifetime measurement, are compared to measure oscillation probabilities as a function of reconstructed neutrino energy. These uncertainties make determining the CP-violating phase, which has yet to be constrained, a serious task.
Precise CP phase values are important not only for confirming the universe's baryon asymmetry but also for describing phenomena outside the standard model, such as the effective mass of neutrinos \cite{jd_eff}.

Neutrinos are large (very small mass) weakly interacting particles that only interact with matter very seldom. It is quite difficult to detect them in regular detectors due to their weakly interacting nature. In comparison to other baseline neutrino experiment detectors, the DUNE collaboration's proposed detector is an exceedingly advanced detector of this generation. This versatile DUNE detector will be built in a modular format, with each module containing unique technologies capable of detecting neutrino occurrences and other unusual phenomena. The charged current (CC) and neutral current (NC) processes interact with neutrinos. In the final stages, an interactive event with the CC process created leptons ($e^{+},e^{-},\mu^{+}, \mu^{-}, \tau^{+}, \tau^{-}$) corresponding to neutrino flavors ($ \nu_{e}, \bar{\nu_{e}},\nu_{\mu}, \bar{\nu_{\mu}}, \nu_{\tau}, \bar{\nu_{\tau}}$) and hadrons.  In our Monte Carlo (MC) sample, interaction events with the NC create the same neutrino flavor and hadrons in the final states. The primary CC processes include CC-Quasi-Elastic (QE), Resonance (RES), Deep Inelastic Scattering (DIS), and Meson Exchange Current (MEC), which are all modeled based on neutrino energy. The cross-sections of these CC interaction mechanisms are energy-dependent.
The investigation of these neutrino-nucleus interactions in the nuclear medium become more difficult. The neutrinos interact with the nucleons inside the target nucleus to form secondary leptons and hadrons in these scattering processes. Because these nucleons are not non-interacting, nuclear effects must be taken into account when researching the physics of neutrino-nucleon interactions. To boost interaction rates, recent neutrino experiments have used  material with a high atomic number as detector materials, which has increased systematic uncertainty in neutrino  reconstruction energy due to nuclear effects. Uncertainties from the multi-nuclear correlation, nuclear Fermi motion effects, binding energy,  and final-state interactions (FSI) of created hadrons in distinct interaction channels are all major nuclear effects present in nuclear targets. At both the momentum
transferred distribution and the double differential cross-sections, the fermionic motion of the nucleons, as well as the
Pauli blocking mechanism, have distinct and noticeable impacts. Furthermore, it was discovered that the nucleon
separation energy has a significant impact on the momentum transferred distribution \cite{perez}. The need for precise neutrino oscillation parameter knowledge is influenced by several factors, the most important of which is the correct reconstruction of neutrino energy. The likelihood of neutrino oscillation is based on neutrino energy, and any inaccuracy in neutrino energy measurement will affect neutrino oscillation parameter measurements because it generates uncertainty in cross-section measurement and event identification. The electron scattering (e-A) data or a simple target like neutrino-hydrogen scattering ($\nu(\bar{\nu})$-H) data can be used to constrain these kinds of systematic errors. We are running a simulation using the latter option to see how many systematic uncertainties  there are in the measurable energy when utilizing the Ar target.

Aside from the systematic uncertainties in the observables due to uncertainty in total neutrino cross-sections \cite{2_c}, the secondary particles that emerge from the nucleus after the final FSI also contribute to systematic inaccuracies \cite{2_cc, 2_d}. One of the key reasons is that the nucleons in the target nucleus are neither at rest nor non-interacting. Recent experiments with intense neutrino beams have resulted in a significant gain in statistics and a reduction in statistical uncertainty. These tests now emphasize the importance of managing systematic uncertainty with caution. Neutrino experiments of this generation use a complex nuclear target for interaction, resulting in significant and non-negligible nuclear effects. When modeling neutrino scattering events, most neutrino experiments use the Impulse Approximation \cite{5a} or the Fermi Gas Model \cite{5b} to account for nuclear effects. Following the results of several neutrino tests conducted in various energy ranges, recent experiments have shown that nuclear impacts must be properly considered. Neutrino experiments like MiniBooNE \cite{2e} and MicroBooNE \cite{3d} that used lower energy neutrino beams are only sensitive to two types of neutrino interactions: QE and RES.

\begin{figure} 
\centering\includegraphics[scale=.4]{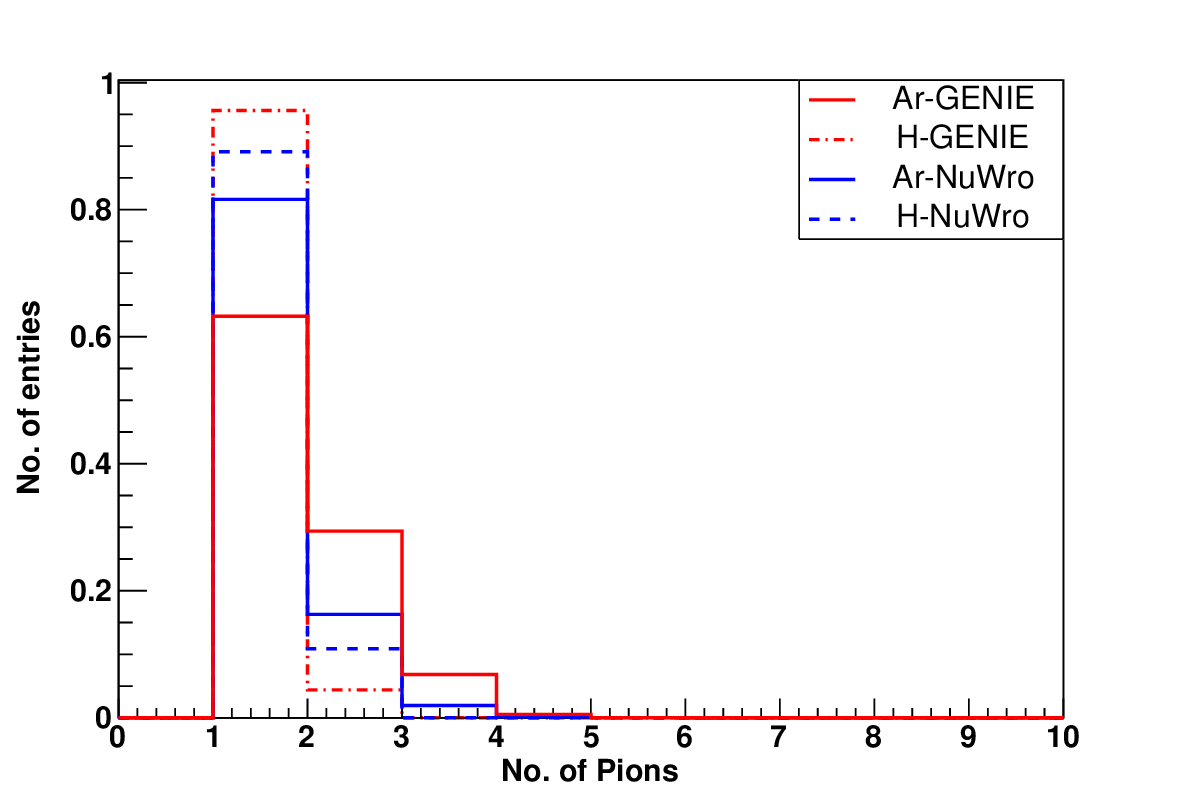}
\centering\includegraphics[scale=.4]{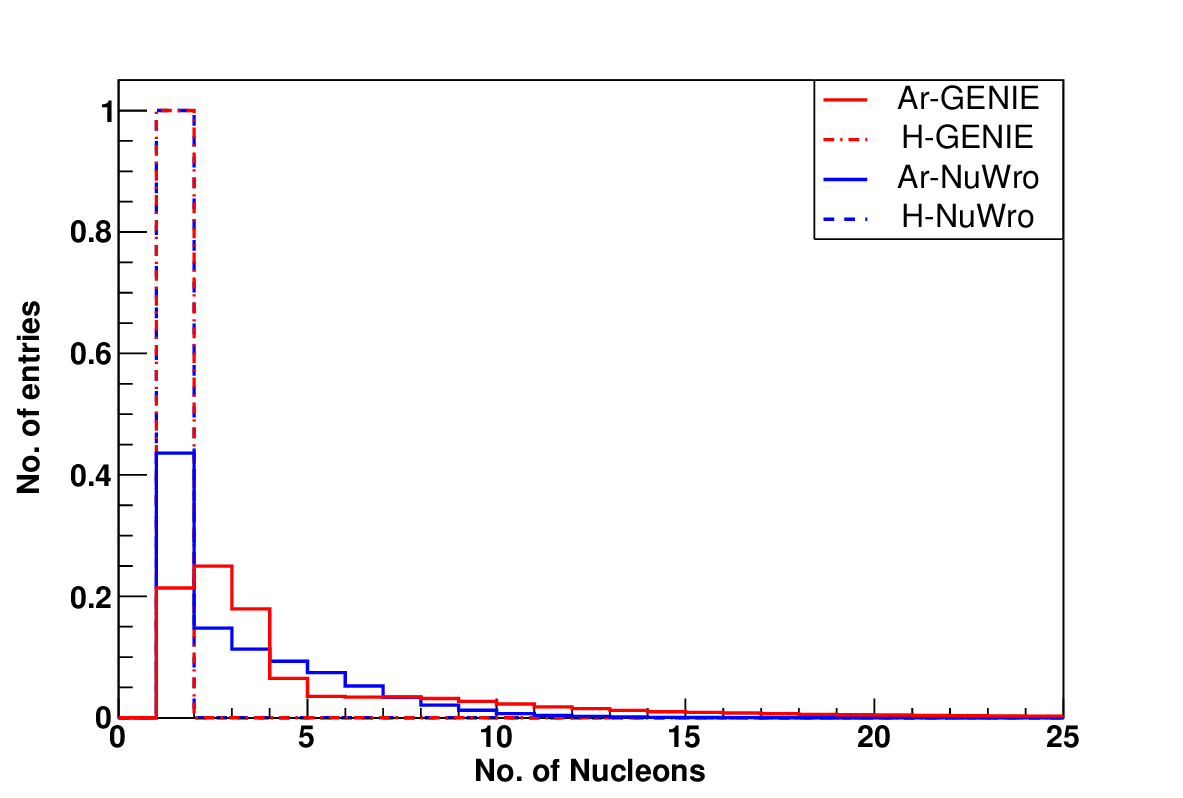}
\caption{Left panel shows the number of pions and the right panel shows the number of nucleons produced in the final states for $\nu_{\mu}$-Ar (solid line)  and  $\nu_{\mu}$-H (dotted line)   interaction. These number of particles are extracted for the RES channels events sample only and interaction is done for the DUNE flux energy range with both the neutrino events generator.  The blue line represents the NuWro sample and  the red line represents the GENIE sample.}
\label{fig:1}
\end{figure}

In these studies, pion generation accounts for only one-third of the overall neutrino-nucleon interaction cross-section. The nuclear binding energy, Fermi motion, and Pauli blocking are key effects that appear mostly at neutrino energies below $\sim$ 0.5 GeV  \cite{vargas}. The pion generation grows to two-thirds of the overall cross-section \cite{5c} as we progress towards higher energy neutrino beam experiments like NOvA, MINERvA, and DUNE. As a result, it is critical to thoroughly quantify the pion production routes under strict quantitative control. The final state interactions of hadrons, which are created in the reaction together with leptons, are one of the most important nuclear processes that cause fake events \cite{jd}. Due to numerous processes such as elastic and charge exchange scattering with the nucleons that exist in the nucleus through strong interactions, pions formed in neutrino-nucleon interactions can be absorbed, change direction, energy, or charge, or even make new pions. As a result, depending on the emerging particles from the nucleus, bogus events occur. In this work, we calculated the contribution of nuclear effects (final state interaction) in resonance channels using neutrino oscillation parameters at DUNE, focusing on the relevance of having a good understanding of pion generation channels in higher energy neutrino beam investigations. Figure \ref{fig:1} shows the number of pions in the left panel and the number of nucleons in the right panel for RES processes for Ar (upper part) and H (lower part) target.

We propose a simulation-based analysis employing two neutrino event generators, GENIE and NuWro, in this paper. We can also perform this analysis by using MARLEY, a novel event generator developed to better satisfy the modeling demands of the low-energy neutrino community (tens of MeV and lower) \cite{marley}. This simulation study is done for the DUNE project, however, similar impacts can be checked for any baseline experiment. The purpose of any baseline experiment is to quantify the neutrino oscillation probability; here, we've done a simulation of what the DUNE baseline and detectors are intended to observe. This analysis is carried out with the target materials Argon (Ar), which will be used in both the ND and FD detectors and then with hydrogen, which will be utilized in the near detector complex to measure the DUNE neutrino flux, with all other parameters remaining the same.
So, we will have two kinds of MC event samples one with Ar target and the second one
with H target. Neutrino energy is reconstructed using a calorimetric method that used all the particles and kinematics
available in the final states. Argon, Iron, and Carbon as a target was extensively used in recent neutrino detectors,
therefore this study will help in the estimation of the systematic uncertainties with the experimental data and model
development. The measurable probability as a function of reconstructed neutrino energy is used to estimate the
systematic uncertainties. The number of systematic inaccuracies in events with the Ar target due to the available nuclear effects in the models used to explain neutrino Ar scattering is approximated as the ratio of probability as a function of reconstructed neutrino energy P(Ar)/P(H).

The following sections comprise the paper: Section \ref{sec2} describes the neutrino event generators GENIE and NuWro employed in this study, as well as a detailed comparison of the nuclear model integrated into them. The flux and oscillation physics are discussed in Section \ref{sec3}. Section \ref{sec4} describes the detector effect and neutrino energy reconstruction then  results and discussion in Section \ref{sec5}. Finally, in Section \ref{sec6}, we provide our conclusions.

\section{EVENT generators: GENIE and NuWro}\label{sec2}

Event generators are used to produce events at detectors so that simulation-based findings can be generated for any experiment to gain an understanding of the predicted observables. Neutrino communities have event generators with built-in models that are modified based on experimental data, and these generators serve as a link between experimental and theoretical physicists. To conduct our research, we chose the two most widely used neutrino event generators: GENIE (Generates Events for Neutrino Interaction Experiments) v3.0.6 \cite{3} and NuWro \cite{4a} version 19.01, which are currently used by the majority of neutrino experiments in the United States and other countries. Many neutrino baseline experiments across the world employ GENIE, including Minerva \cite{3b}, MINOS \cite{3c}, MicroBooNE \cite{3d}, NoVA \cite{3e}, and ArgoNEUT \cite{3g}  experiments. Most neutrino experiments currently employ NuWro as a supplemental neutrino events generator, thanks to a group of physicists led by Cezary Juszczak et al. at Wroclaw University \cite{4a}. GENIE was created with a focus on scattering in the few GeV energy range, which is critical for current and future oscillation studies experiments, while NuWro covers the neutrino energy scale from 100 MeV to 100 GeV. NuWro's core design is based on well-known MCs such as NEUT \cite{4c} and GENIE \cite{3}. It controls all key neutrino-nucleus interactions, as well as DIS hadronization and the intranuclear cascade. NuWro is a tool for estimating the significance of various theoretical models that are currently being studied. NuWro is built around an event framework that has three types of particle vectors: incoming, transient, and departing. GENIE is a modern and adaptable neutrino event simulation platform. The methods considered for neutrino-nucleus interaction processes in both generators and the general technique employed for dealing with neutrino-nucleus interaction analyses. The Relativistic Fermi gas (RFG) model is used in GENIE, which is based on the model proposed by A. Bodek and J.L. Ritche \cite{3h}, but the Local Fermi gas (LFG) model is used in NuWro. In both generators, the Llewellyn Smith model \cite{3i} is used to model QE scattering.
GENIE and NuWro use the newest BBBA07 \cite{3n} and BBBA05 \cite{3j} vector form factors, respectively. GENIE employs a variable axial mass $M_A$ of 0.99-1.2 $GeV/c^{2}$, whereas NuWro employs a variable axial mass of 0.94-1.03 $GeV/c^{2}$. We employed the axial mass $M_A$=1 $GeV/c^{2}$ in our research. In GENIE,  $\Delta$ contribution and other resonance modes were examined separately using the Rein-Sehgal model \cite{3k}, but in NuWro, we only have $\Delta$ contribution and the rest of the resonance modes are averaged using the Adler-Rarita-Schwinger model \cite{3o} for RES events. NuWro uses the Quark-Parton model \cite{3p} for DIS events, while GENIE uses the Bodek and Yang model \cite{3l}.
 
 \section{  DUNE flux and oscillation probability }\label{sec3}
 
\begin{figure}
\centering\includegraphics[scale=.4]{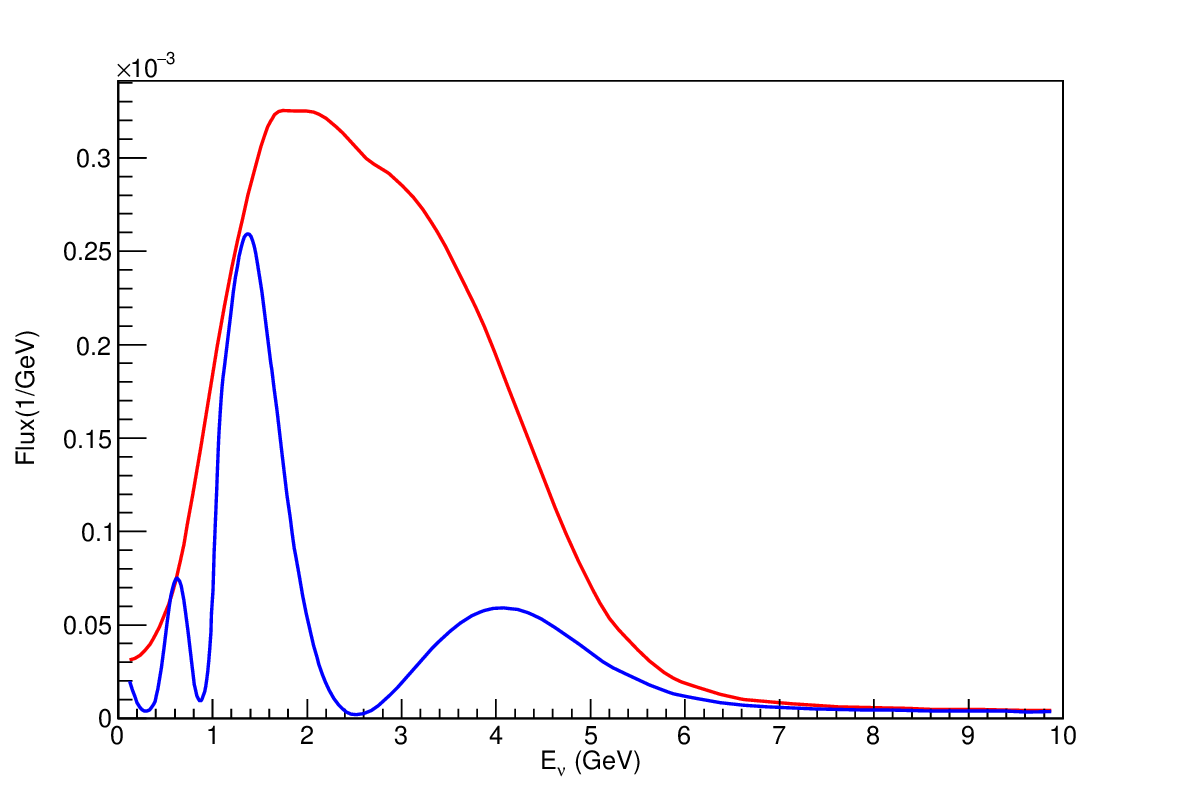}
\caption{The available neutrino ($\nu_{\mu}$) flux at DUNE ND and FD site as a function of true neutrino energy.}
\label{fig2}
\end{figure}
 We have generated an inclusive MC sample of 1 million events at the ND site using this flux for each target (Ar and H) for further performing the simulation and analysis, which  will be discussed in the following sections. This flux is also folded with the oscillation probability to get the flux at the FD site shown in  Figure \ref{fig2}, again the folded flux is used to get the similar at the FD site. The oscillation probability used here is represented by equation \ref{eq1} and the true set of values of the oscillation parameters \cite{2_b} considered in this analysis are presented in  Table \ref{table1}.
\begin{table}[htp]
	\caption{Oscillation parameters considered in our work.}
	\renewcommand\thetable{\Roman{table}}
	\centering
	\setlength{\tabcolsep}{2pt}
	\begin{tabular}{c | c  }
\hline\hline
\textbf{Parameter}                        &  \textbf{Central Value}      \\
\hline
 $\theta_{12}$(NO)           &   $0.5903$                                     \\
$\theta_{13}$ (NO)              &   $0.150$                            \\
$\theta_{23}$(NO)           &   $0.866$         \\
$\delta_{CP}$               &   $90^{\circ}$          \\
$\Delta m^{2}_{21}$ (NO)        &   7.39e-5$eV^{2}$                         \\
$\Delta m^{2}_{32}$ (NO)    &  2.451e-3 $eV^{2}$     \\
$\Delta m^{2}_{31}$(NO)     &  2.525e-3  $eV^{2}$     \\
$\rho$                      &  2.848 $g$ $cm^{-3}$    \\
\hline\hline

\end{tabular}
\label{table1}
\end{table} 
One can see that the oscillation parameters and neutrino energy are involved in the oscillation probabilities equation, the uncertainties in the measured neutrino energy directly translate into uncertainties in oscillation parameters. The oscillation probability of $\nu_{\mu}\rightarrow \nu_{e}$ channel through matter for the three-flavor scenarios and constant density approximation is written as, \cite{2_b}

\begin{multline}	\label{eq1}
P(\nu_{\mu}\rightarrow\nu_{e})=  \sin^2\theta_{23}\sin^22\theta_{13}\dfrac{\sin^2 (\Delta_{31} -aL )}{(\Delta_{31} -aL )^2}\Delta_{31}^2 + \sin 2\theta_{23} \sin 2\theta_{13} \sin 2\theta_{12} \dfrac{\sin(\Delta_{31} -aL )}{(\Delta_{31} -aL )} \Delta_{31} \\
 \times \dfrac{\sin(aL)}{aL} \Delta_{21} \cos(\Delta_{31}  +\delta_{CP})+\cos ^2 \theta_{23} \sin^2 2\theta_{12} \dfrac{\sin^2aL}{(aL)^2} \Delta_{21},
\end{multline}

where
\begin{equation}\label{eq2}
	a= \dfrac{G_F N_e}{\sqrt{2}} \approx \dfrac{1}{3500 \space km}\left(\dfrac{\rho}{ {3.0} {g/cm^3}}\right) ,
\end{equation}

$G_F$ is the Fermi constant, $N_e$ denotes the number density of electrons in the Earth\textsc{\char13}s crust, $\Delta_{ij}=$ 1.267 $\Delta m^2_{ij}L/E_\nu$, $L=1285$ is the distance from the neutrino source to the detector in km, and $E_{\nu}$ stands for neutrino energy in GeV.

\section{ Detector effects and energy reconstruction}\label{sec4}

DUNE at Long-Baseline Neutrino Facility (LBNF)  will be consist of ND and FD with similar target material but the difference in dimension and technology. The ND system will be located at Fermilab, 574 m  downstream and 60 m underground \cite{2_b} from the neutrino beam source point. The baseline design of the DUNE ND \cite{D1} system comprises of three main components: (1) A 50t LArTPC called ArgonCube with  pixellated readout; (2) A multi-purpose tracker, the high-pressure gaseous argon TPC (HPgTPC), kept in a 0.5T magnetic field and surrounded by electromagnetic calorimeter (ECAL), together called the multi-purpose detector (MPD); (3) and an on-axis beam monitor called System for on-Axis Neutrino Detection (SAND). SAND \cite{1DUNE} will consist of  an Electromagnetic Calorimeter (ECAL), a superconducting solenoid magnet, a thin active Lar target, a 3D scintillator tracker (3DST) as an active neutrino target, and a Low-density tracker to precisely measure particles escaping from the scintillator. The FD (and the similar ND) has to be made of heavy nuclei rather than hydrogen is  the main  source of complication.  The primary goal of SAND is constraining systematic uncertainties from nuclear effects.
For a better understanding of nuclear effects in neutrino-nucleus interactions, it is useful to examine  what would happen if the detectors were made of hydrogen because in a detector made of hydrogen, the initial state is a proton  at rest and there is no FSI. The 40 kt DUNE FD  will comprise  of four separate LArTPC detector modules, each with a fiducial volume (FV) of at least 10 kt, installed $\sim $1.5 km underground at the SURF \cite{D2}. The internal dimensions  15.1 m (w)$\times$14.0 m (h)$\times$62.0 m (l) of each LArTPC fits inside an FV containing a total LAr mass of about 17.5 kt. Detailed detector simulation and reconstruction analysis is done by the DUNE-collaboration that can be found in detail in reference \cite{2_b, D3}, we have applied the same detector efficiency and resolution for the particles available in the FSI while reconstructing the neutrino energy in our analysis.
 
\begin{figure}[ht!]
\centering\includegraphics[scale=.4]{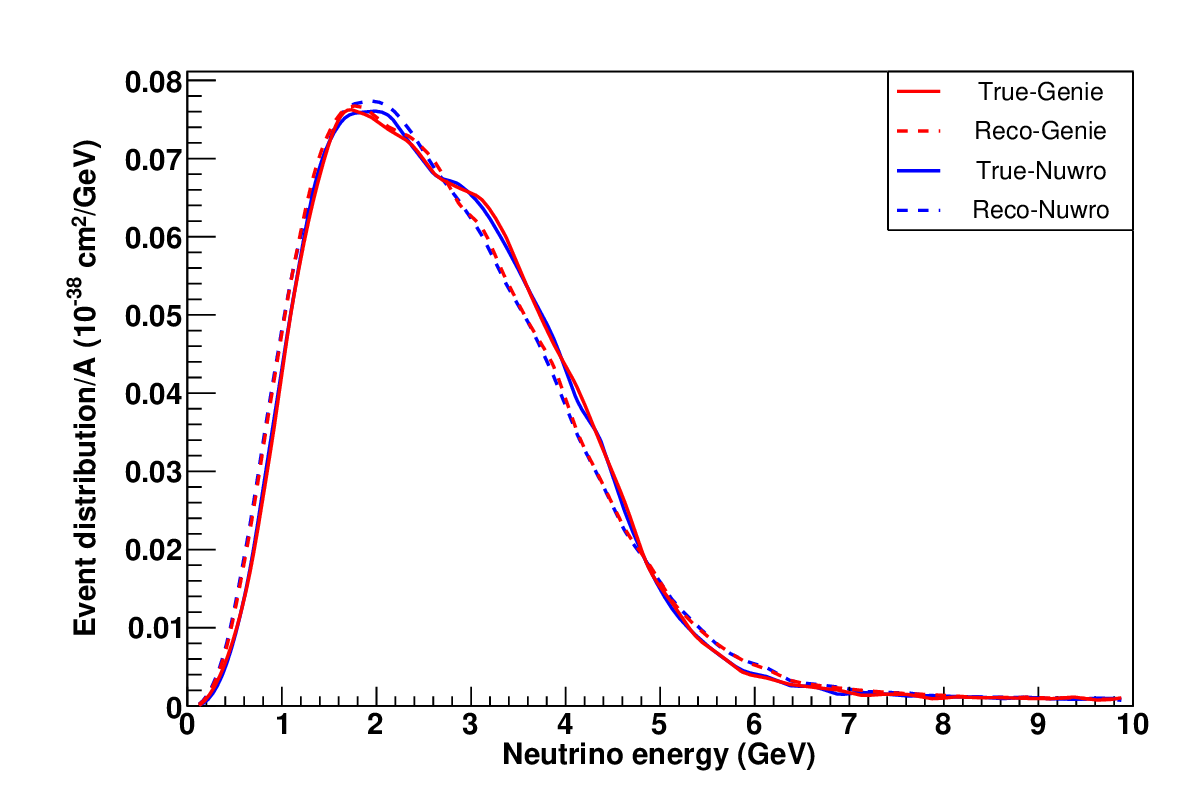}\centering\includegraphics[scale=.4]{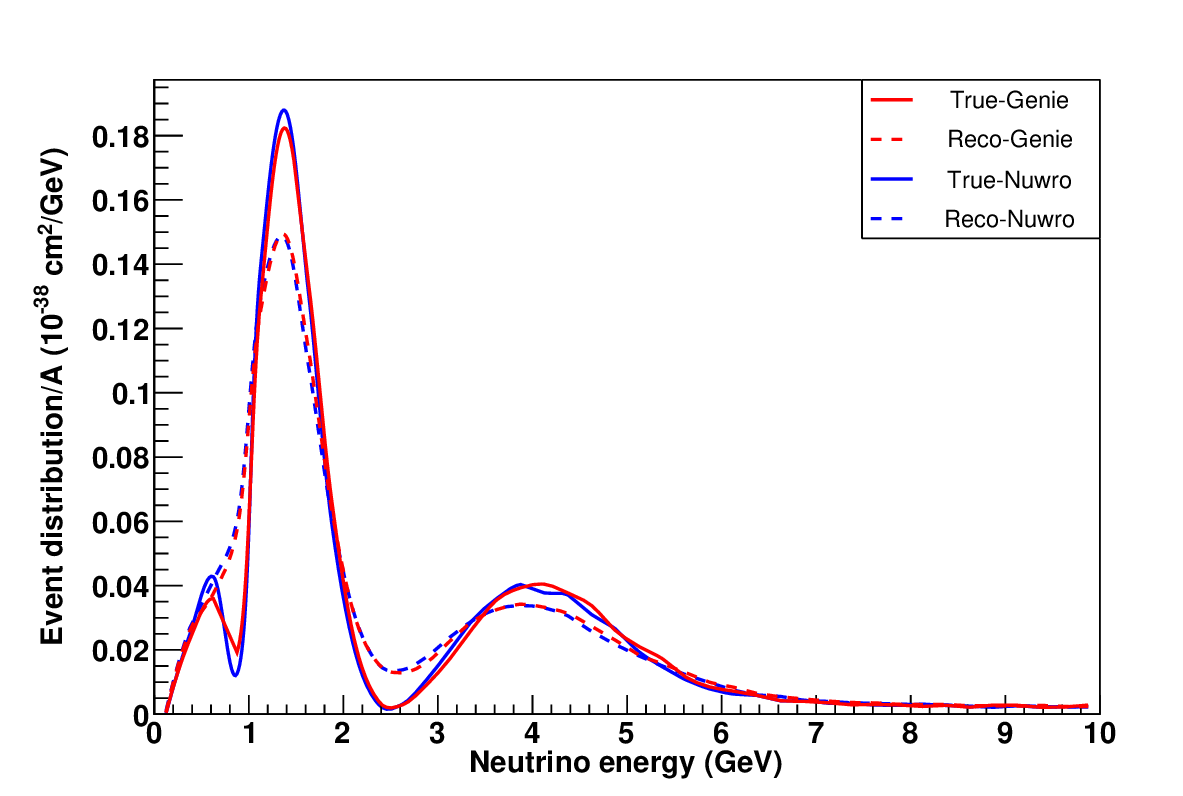}
\caption{Left and right panel shows the QE events sample as a function of true (solid lines) and reconstructed (dashed line) neutrino energy at ND and FD sites respectively  for $\nu_{\mu}$. The red line event rate is obtained with GENIE, while the blue line with NuWro, we will follow the same color and line scheme in the following sections. }
\label{fig3}
\end{figure}
\begin{figure}[ht!]
\centering\includegraphics[scale=.4]{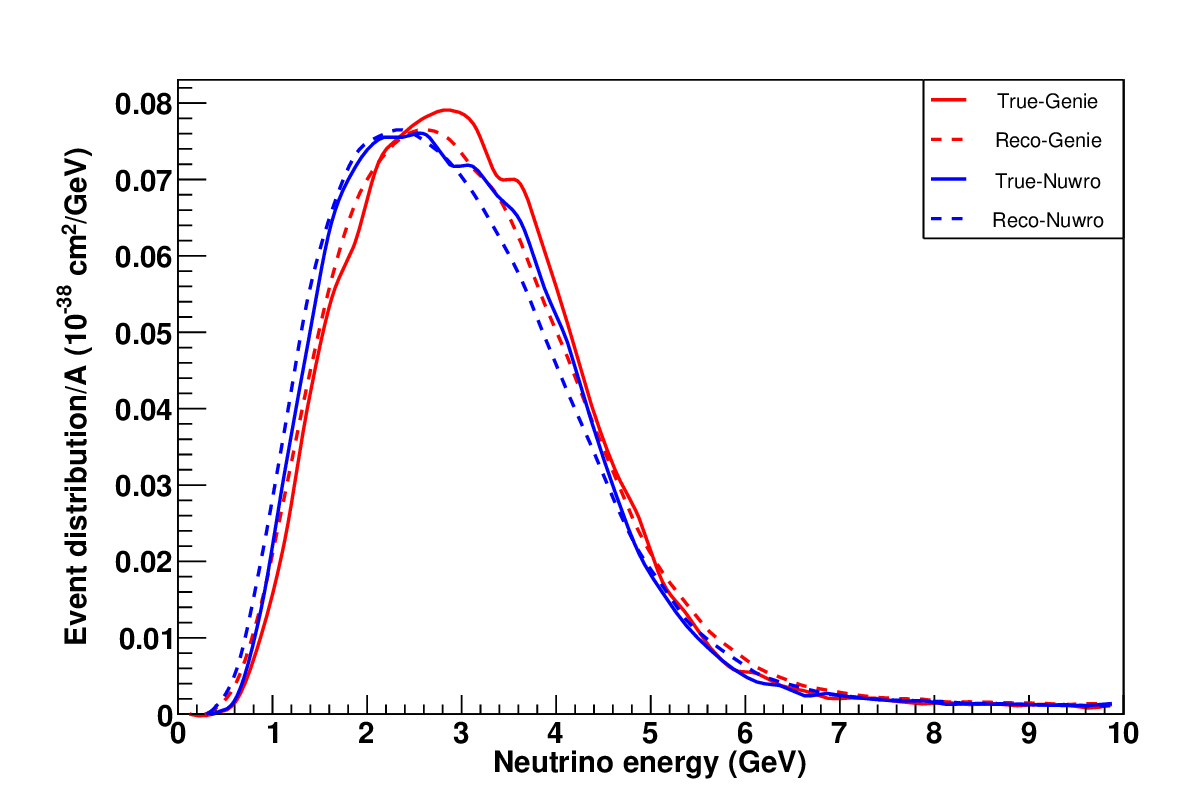}\centering\includegraphics[scale=.4]{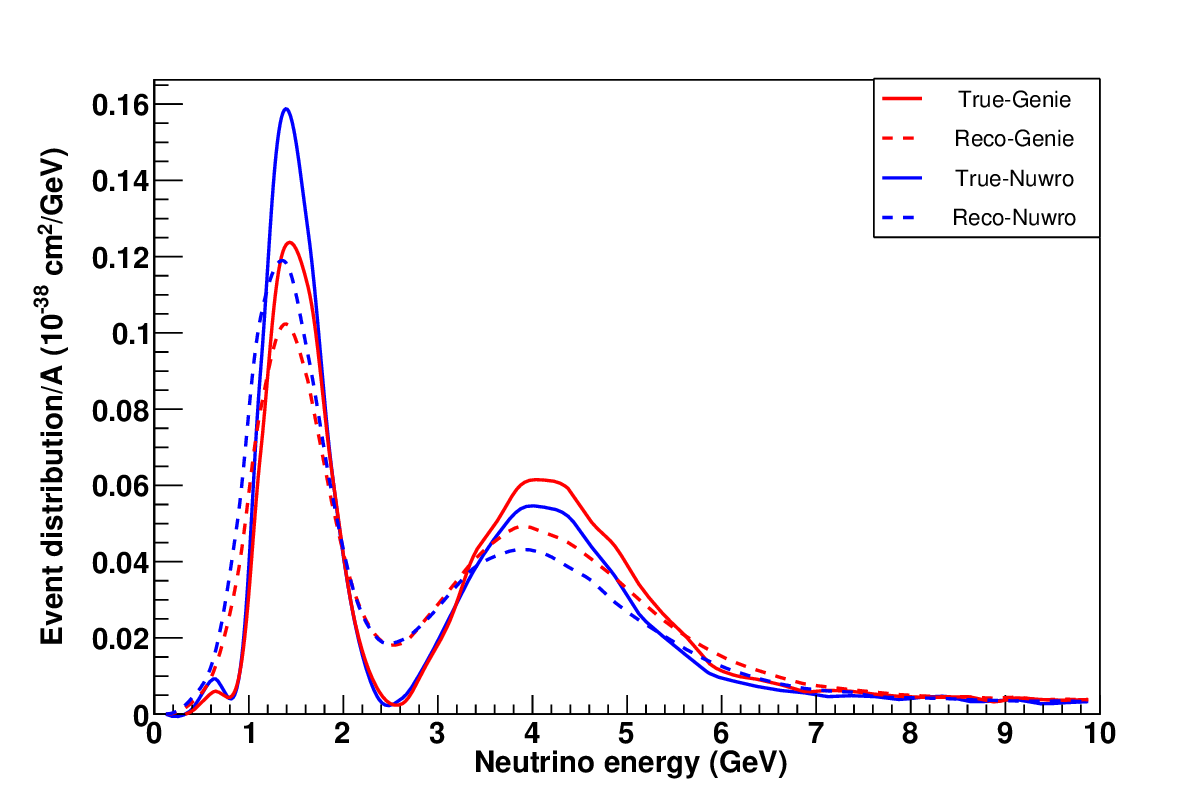}
\caption{Left and right panel shows the RES events sample as a function of true and reconstructed neutrino energy at ND and FD site respectively for $\nu_{\mu}$. The color and line scheme is the same as described in the earlier Figure 3 caption. }
\label{fig4}
\end{figure}
For accurate oscillation analysis, the reconstruction of neutrino energy from the kinematics of the particles available in the FSI is of prime importance. This energy has to be reconstructed on an event-by-event basis because the neutrino beams are generated via secondary decay of primarily produced hadrons which assign a broad range of energies to the neutrinos. We can make use of either of the two schemes for neutrino energy reconstruction  i.e., (1) Kinematic method \cite{2d,2e,2f,2g}; (2) Calorimetric method \cite{2h,2i}. The kinematic method of energy reconstruction is based on the premise that the incoming neutrino interacts with a single nucleon which is assumed to be at rest, bound with constant energy, and the outgoing particles include a single nucleon i.e., no other nucleons are knocked out of the nucleus and a lepton. As the energy of the incoming neutrino increases, the contribution of inelastic processes-RES, and DIS becomes  dominant and  may lead to many hadrons in the final state thus leading to inappropriate estimation of neutrino energy reconstruction. Thus this method is largely used by lower energy experiments, such as MiniBooNE \cite{2e} and T2K \cite{2_cc}. The calorimetric method of neutrino energy reconstruction demands information of all the detectable final state particles on an event-by-event basis i.e., a total of the energy deposited by all reaction products in the detector and thus permitting for precise reconstruction of neutrino energy. It is appropriate for all types of events, unlike the kinematic method which is only applicable for QE event reconstruction. Still, this method gives rise to challenges in the way of true neutrino energy reconstruction. The main challenges are the precise reconstruction of all produced hadrons, where neutrons are assumed to escape detection completely. Thus, the calorimetric method relies on the capability of detector design and performance in reconstructing final states. Nuclear effects also play an important role in incorrect neutrino reconstructed energy as they may promote a significant amount of missing energy, the energy that gets absorbed in the intermediate state interactions before the final states particles come out of the nucleus, obstructing the reconstruction of neutrino energy \cite{2i}. For example, if a meson produced at the initial state is absorbed in the observer system, in general, its energy is not deposited in the calorimeter.  The neutrino reconstructed energy $ E^{Calori}_{\nu} $ in charged-current (CC) processes resulting in the knockout of $n$ nucleons and production of $m$ mesons, can be simply determined using the calorimetric method, 

\begin{equation}\label{eq3}
	E^{Calori}_{\nu} =   E_{lep}+ \epsilon_{nuc}+\sum_{a}(E_{n_{a}}-M)+ \sum_{b}E_{m_{b}}
\end{equation}
Where $E_{lep}$ is the  energy of outgoing lepton, $ E_{n_{a}}$ denotes the energy of the $a^{th}$ knocked out nucleon, $ \epsilon_{nuc} $ denotes the corresponding separation energy  and the single-nucleon separation energy is fixed to 34 MeV \cite{2i}. The same value of $\epsilon_{nuc}$ is added in the calorimetric energy reconstruction for every nucleon detected. $E_{m_{b}}$ stands for the energy of $b^{th}$ produced meson, and $M$ is the mass of nucleon. In our analysis, we have used calorimetric method for energy reconstruction. Figure \ref{fig3} shows the event rates as a function of true and reconstructed neutrino energy for QE with 0 pions, exactly 1 proton and X neutrons \cite{mosel} events both at  ND, without oscillations in the left panel, and at the FD, with oscillations, in the muon disappearance channel in the right panel by considering the normal hierarchy and $\delta_{CP}$=$\pi/2$ using GENIE and NuWro generator. Figure \ref{fig4} shows the  event rates as a function of true and reconstructed neutrino energy for RES processes  for $\nu_{\mu}$. Events distribution for the fixed neutrino energy 2.5 GeV is also shown in Appendix Figure \ref{fig:8} and Figure \ref{fig:9} for QE and RES process respectively, this is the mean neutrino energy at which DUNE flux will be peaking.  

\begin{figure}
\centering\includegraphics[scale=.4]{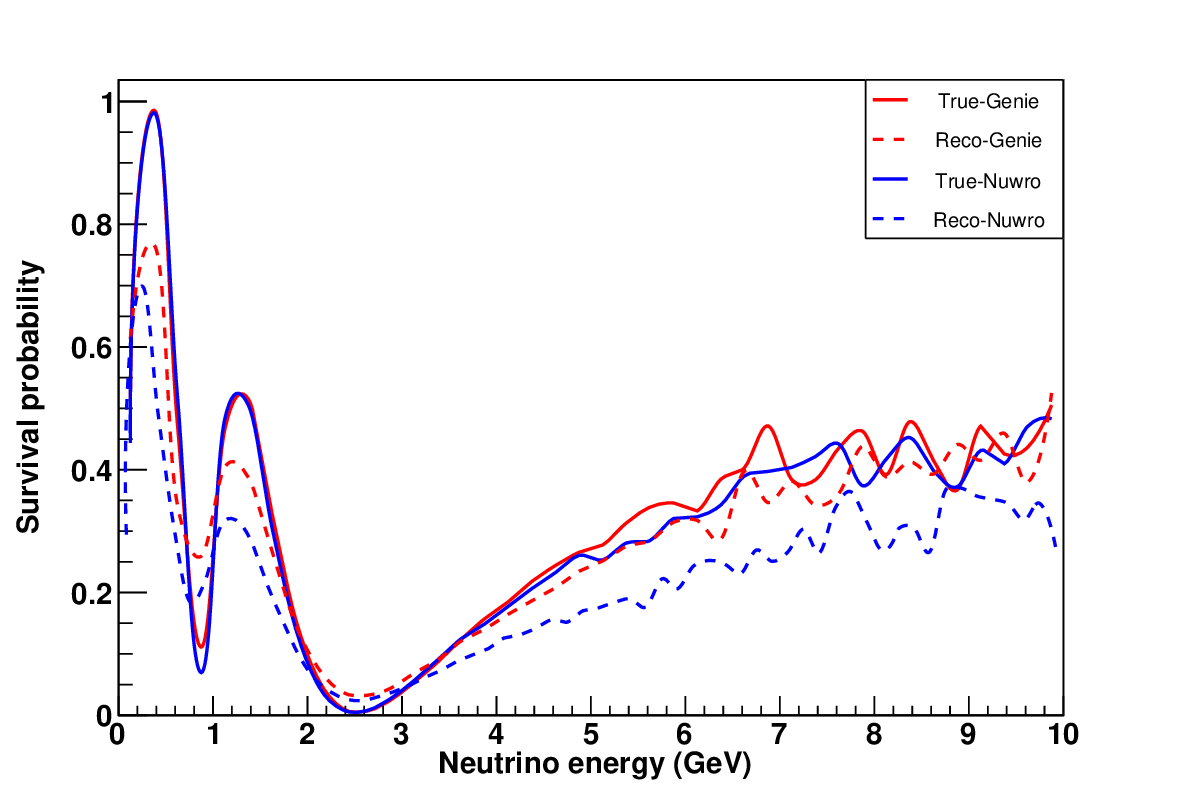}\centering\includegraphics[scale=.4]{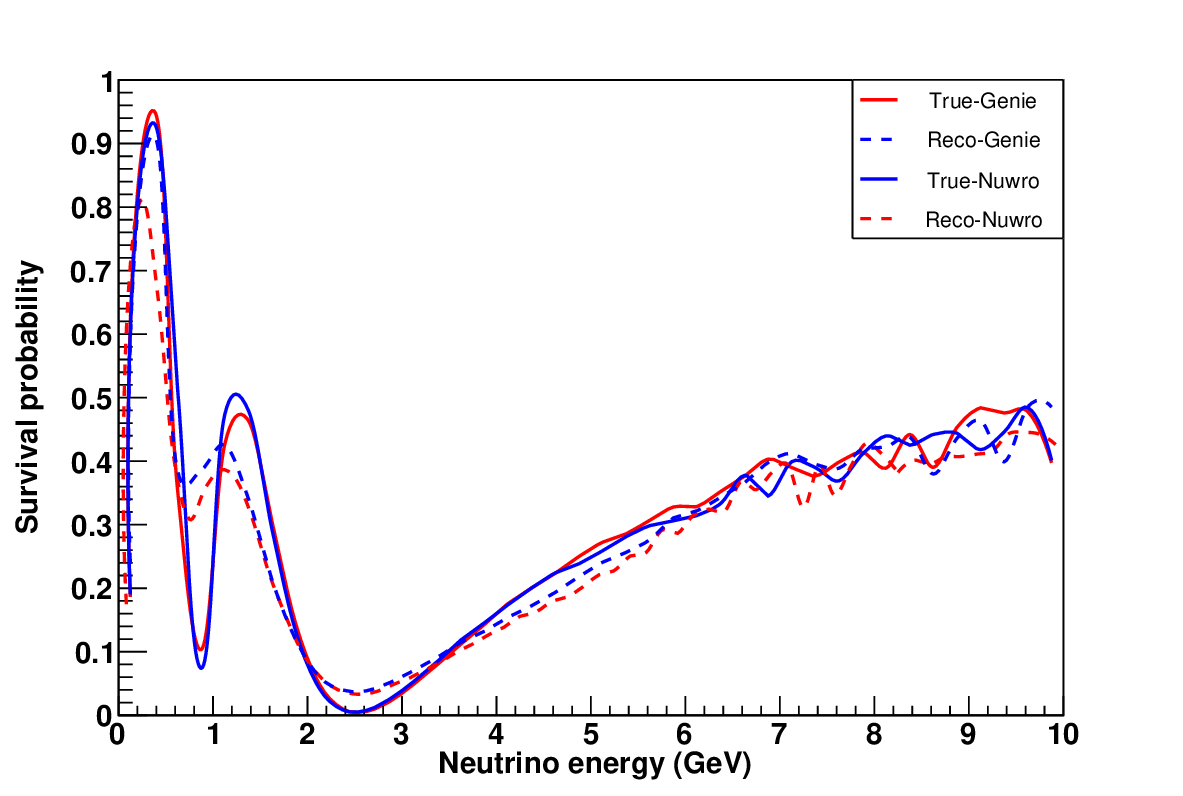}
\caption{Survival probability for $\mu$-neutrinos both for true (solid) and reconstructed (dashed) energies for QE (left)  and RES (right) processes  for Ar target. The (red) curves give the probability in GENIE, (blue) curves give the probability in NuWro.}
\label{fig5}
\end{figure}

\begin{figure}
\centering\includegraphics[scale=.4]{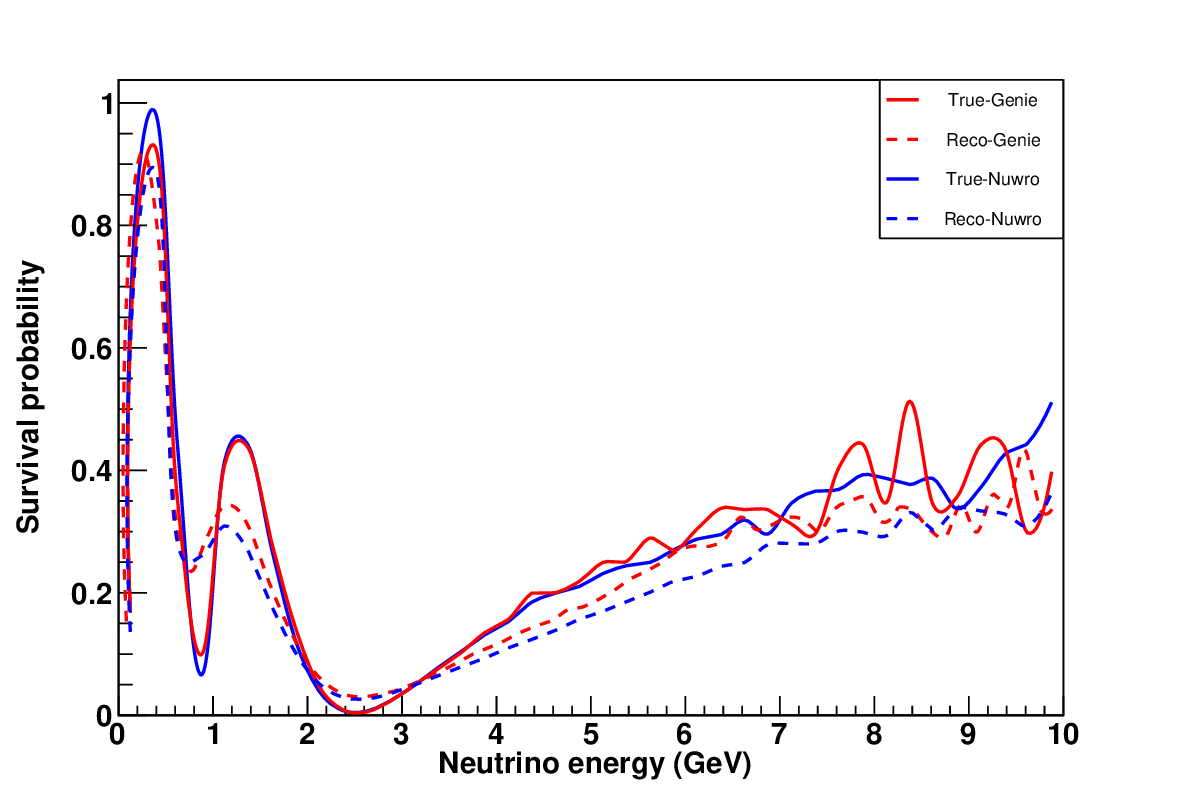}
\caption{Survival probability for  $\mu$-neutrinos  for H target for RES processes as a function of reconstructed (dashed curve) and true (solid curve)  neutrino energy. }
\label{fig6}
\end{figure}

\section{Results and discussion}\label{sec5}
We should have a good understanding of the hadronic physics of neutrino-nucleus interactions since it is critical to understand nuclear effects. The information about the energy dependence of all exclusive cross-sections and nuclear effects is important for the construction of a nuclear model. Any ignorance of theoretical uncertainties in the nuclear models while simulating results cost inaccuracy. We use different event generators, which are built upon these nuclear models to test results from the neutrino oscillation experiments. These predict the neutrino-nucleus interaction rates for a specific nuclear target along with the topology of final state particles which is crucial for oscillation analysis. With  the use of complex targets, nuclear effects play a vital  role in the prediction of neutrino oscillation physics. Different generators use different approximations to measure the nuclear effects giving rise to different results. Our analysis of nuclear effects in this work reflects the extent of uncertainty in the physics of the selected neutrino event generators. The QE and RES interaction channels are analyzed in this article.

From  Figure \ref{fig:1}, we observe that the average number of pions and nucleons in the final states are significantly different in both generators. The understanding of this difference in number is more complicated for the bigger target materials (Ar) compared to the smaller target materials (H). The actually produced mesons or another secondary at the vertex into the nuclear environment are easily get absorbed or have further interactions within the nucleus. This leads to the misidentification of the events channel process from the original channels process produced. This difference can not be underestimated and it reflects the lack of understanding of the theoretical models or nuclear physics that are recently being used in generators to describe the final states interactions and their  hadronization. This is the major systematic error available in the theoretical model that causes the wrongly reconstructed neutrino energy. Due to this incomplete understanding of theoretical  models available in the generators that are being used in experiments to tune the data are not sufficient to explain the scattering and cross-sections. Many theoretical physics groups have tried to define the nuclear medium in the different neutrino event generators to overcome this problem but more physics input is required to attend to this problem properly. The miss-reconstructed  neutrino energy is the major cause of systematic uncertainties in cross-sections measurements at the neutrino experiments. The theoretical cross-section models for the energy range of flux planned at the DUNE project with the NuMI (Neutrinos at the Main Injector) beam is still not very well developed and must be well developed to achieve the kind of accuracy is needed to be achieved the required goal. This work will help to get some  systematic uncertainties available in the models. The DUNE neutrino beam below 1 GeV, QE scattering is the main mode of neutrino-nucleon interaction but as we increase energy, the RES and DIS processes rule the QE interactions. As we see neutrino beam peaks around 2.5 GeV and in this energy region RES production is the dominant which mainly results in the production of pions. In  Figure \ref{fig3} and Figure \ref{fig4}, where total event distribution is shown as a function of true and reconstructed energy for QE and RES processes respectively for $\nu_{\mu}$, we can see for QE processes, there is not much difference but for RES processes there is little difference for energy from 0.125-1.5 GeV and  3-7 GeV in both generators.

\begin{figure}
\centering
\includegraphics[scale=.4]{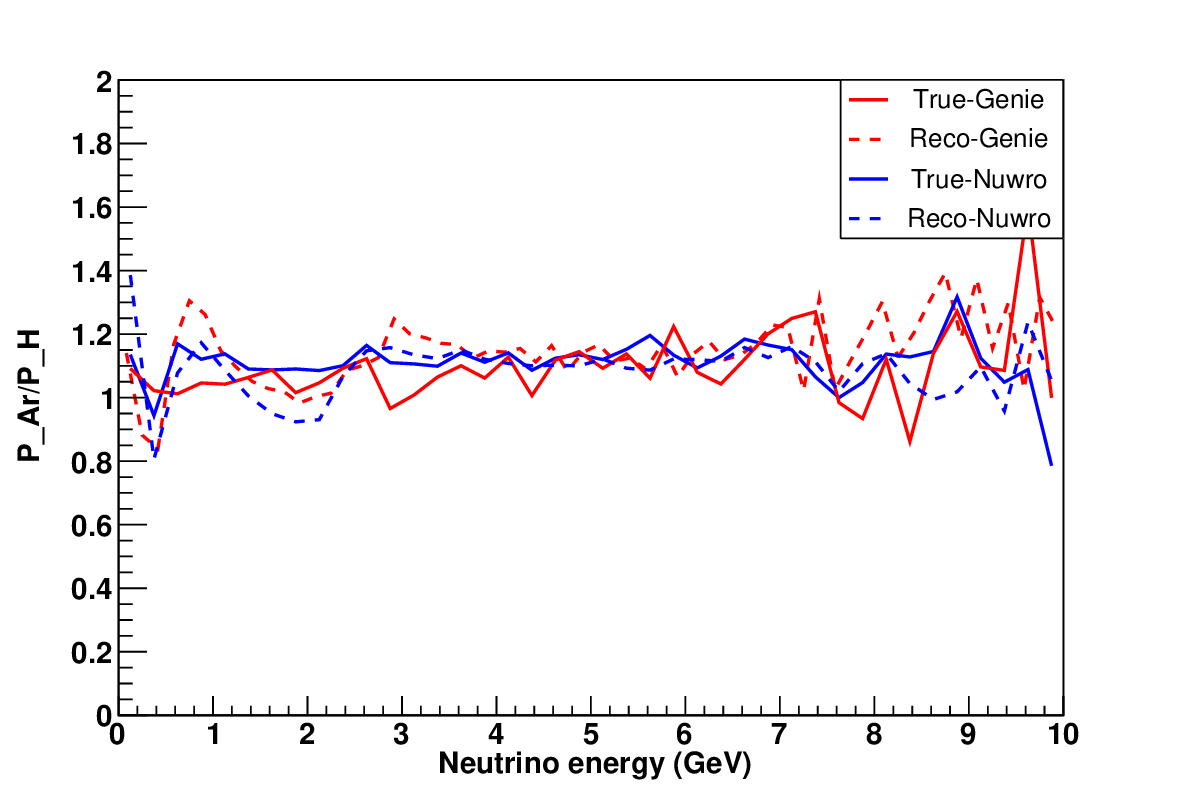}
\caption{Ratio of oscillation probability with Argon to Hydrogen target as a function of reconstructed (dashed) and true (solid) neutrino energy for RES events. The (red) curve for GENIE and (blue) for NuWro.}
\label{fig7}
\end{figure}
One can extrapolate the oscillation probability by taking the ratio of events at FD to the events at ND \cite{mosel}. The result is shown in Figure \ref{fig5}, both as a function of true and reconstructed neutrino energy for QE processes (left) and  RES processes (right) for Ar target. For quantifying the nuclear effects in the Ar (A=40) target, which is going to be used in the DUNE detector, we use a small target like H which will be used in the SAND detector that has the least nuclear effects since it has only one nucleon. Figure \ref{fig6} shows the survival probability for H target for the RES channel  as a function of reconstruction  and true neutrino energy. The ratio of the probability distribution of Ar/H is shown in Figure \ref{fig7} which gives us an idea about  nuclear effects are available in the Ar target concerning the  Hydrogen for RES events. In the Figure, we can see  that there is a deviation from the unity which means large systematic uncertainty in the generators.

\section{Conclusions}\label{sec6}
We should have a good understanding of the hadronic physics of neutrino-nucleus interactions since it is critical to
understand nuclear effects. The information about the energy dependence of all exclusive cross-sections and nuclear
effects is important for the construction of a nuclear model. Any ignorance of theoretical uncertainties in the nuclear
models while simulating results cost inaccuracy. We use different event generators, which are built upon these nuclear
models to test results from the neutrino oscillation experiments. These predict the neutrino-nucleon interaction rates
for a specific nuclear target along with the topology of final state particles which is crucial for oscillation analysis.
With the use of complex targets, nuclear effects play a vital role in the prediction of neutrino oscillation physics.
Different generators use different approximations to measure the nuclear effects giving rise to different results. Our
analysis of nuclear effects in this work reflects the extent of uncertainty in the physics of the selected neutrino event
generators. The QE and RES interaction channels are analyzed in this article.

\section*{Acknowledgment} 
One of the authors, Miss Ritu Devi offers most sincere gratitude to the Council of Scientific and Industrial Research (CSIR), Government of India, for the financial support in the form of Senior Research Fellowship, file no. 09/100(0205)/2018-EMR-I. 

\section*{Declarations}
\textbf{Conflicts of interests} The authors declare no potential conflict of interests.

\newpage
\appendix
\section{Appendix }

\begin{figure}[ht!]

\centering\includegraphics[scale=.4]{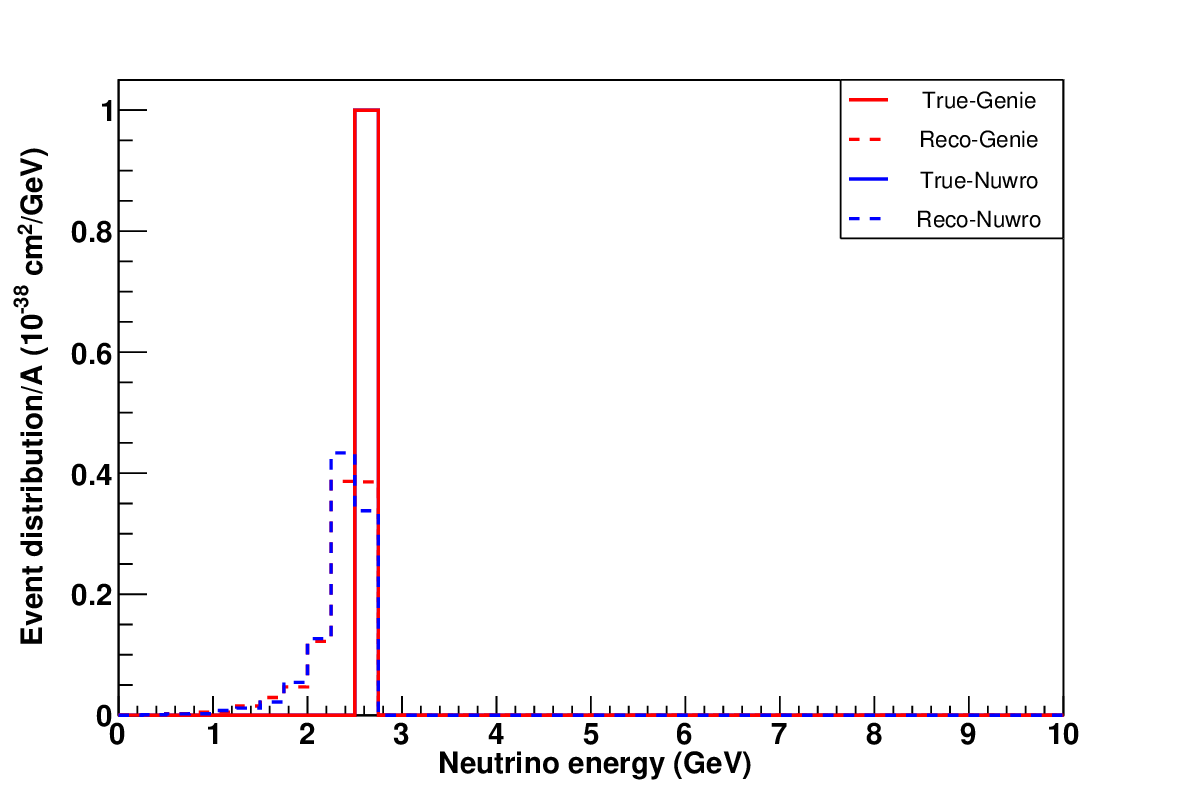}
\caption{QE events sample as a function of true (solid lines) and reconstructed (dashed line) neutrino energy  at 2.5 GeV energy for Ar target.}
\label{fig:8}
\centering\includegraphics[scale=.4]{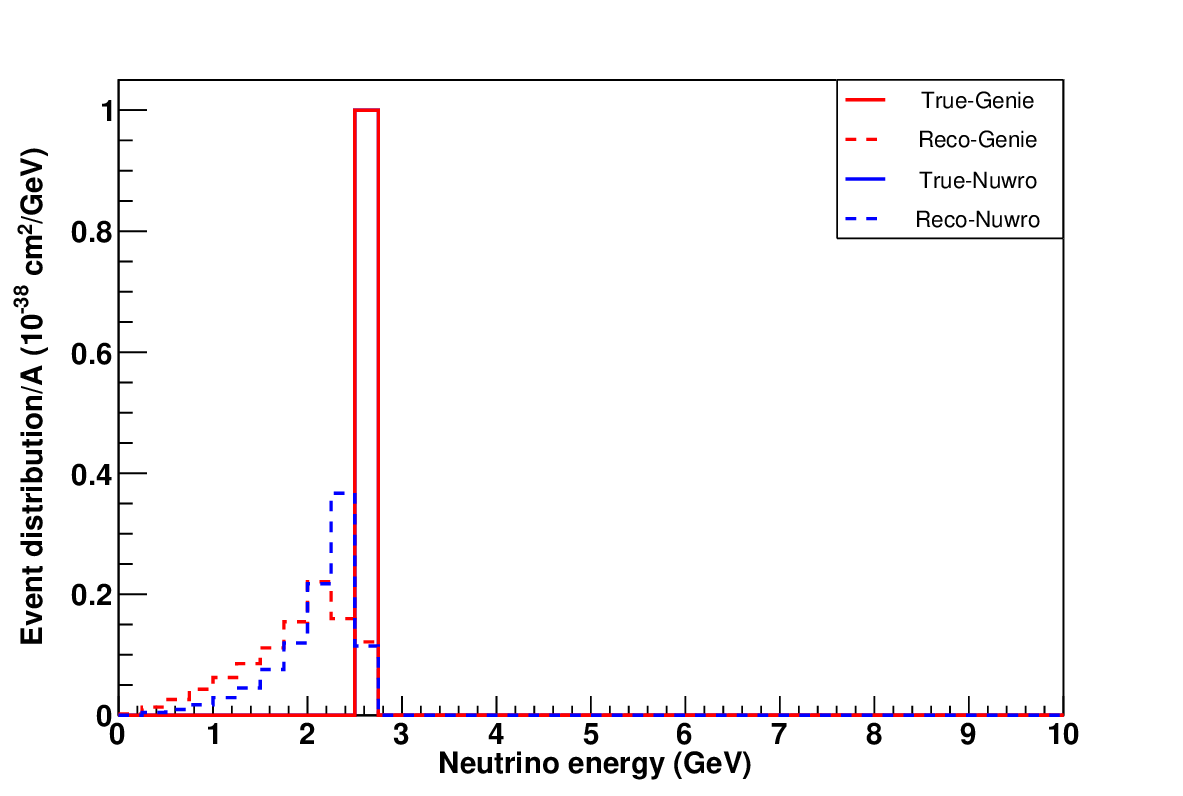}\centering\includegraphics[scale=.4]{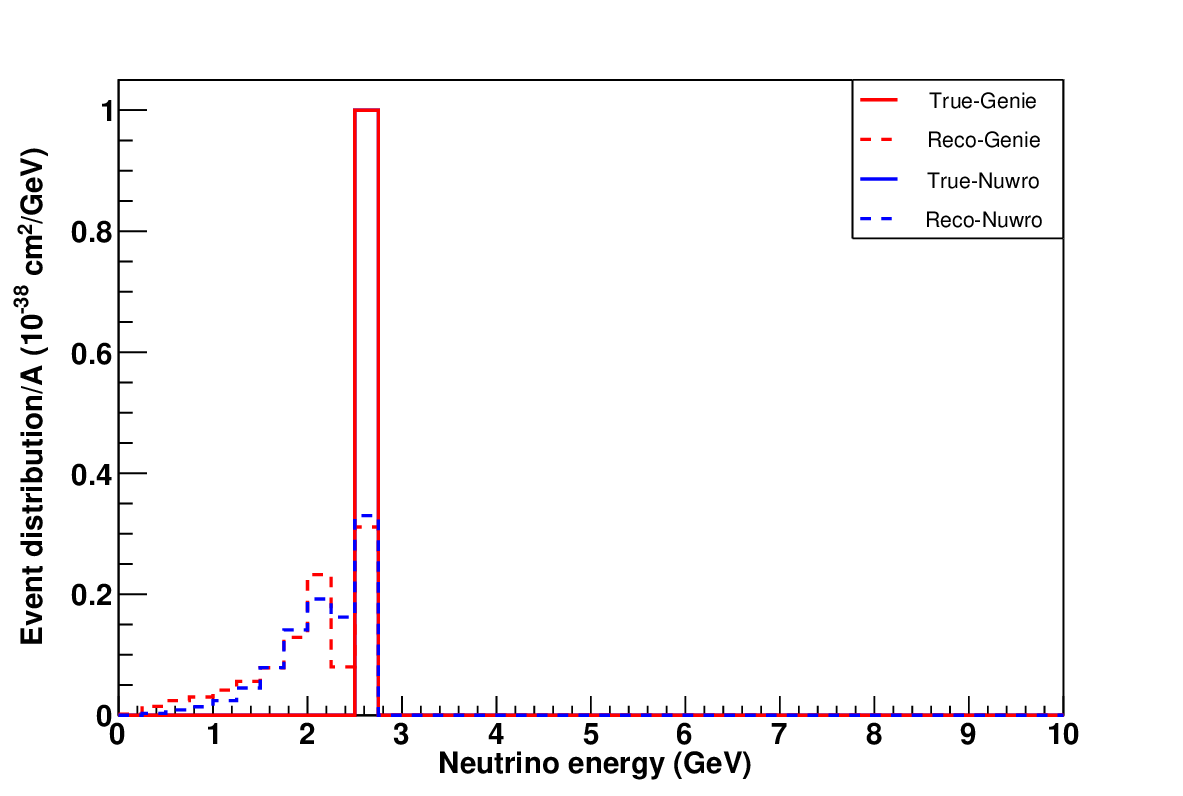}
\caption{RES events sample as a function of true (solid lines) and reconstructed (dashed line) neutrino energy  at 2.5 GeV energy for Ar (left)  and Hydrogen (right)target.}
\label{fig:9}
\end{figure}

\end{document}